# An 860 MHz Wireless Surface Acoustic Wave Sensor with a Metal-Organic Framework Sensing Layer for $CO_2$ and $CH_4$

Jagannath Devkota, David W. Greve, Tao Hong, Ki-Joong Kim, Paul R. Ohodnicki

***Abstract*—Wireless and passive surface acoustic wave (SAW) devices with nanoporous metal-organic framework (MOF) sensing layers are attractive gas sensors for applications in many fields such as energy industries and air pollution control. Here, we report on enhancing the sensitivity and detection limit of zeolitic imidazolate framework-8 (ZIF-8) MOF-coated SAW reflective delay line mass sensors by increasing the operating frequency for sensitive detection of carbon dioxide ($CO_2$) and methane ($CH_4$) at ambient conditions. In particular, we show at least four times higher sensitivity of an 860 MHz (4-µm periodicity) SAW reflective delay line coated with a 240 nm thick ZIF-8 compared to the sensitivity of a 430 MHz (8-µm periodicity) otherwise identical sensor device to the targeted gases. The detection limits of the higher frequency wireless devices for $CO_2$ and $CH_4$ were estimated to be 0.91 vol-% and 7.01 vol-%, respectively. The enhanced sensitivity for higher frequency devices is explained in terms of the frequency dependent acoustic wave energy confinement.**

## I. INTRODUCTION

WIRELESS surface acoustic wave (SAW) devices are potential sensors for monitoring environmental trace gases amongst other chemicals without the need for a dedicated power supply [1-3]. In addition, they offer several advantages including high sensitivity, fabrication simplicity, cost efficiency, and small size [2] as well as longer read range compared to other passive wireless devices such as semiconductor integrated circuit-based sensors [4]. SAW device configurations such as resonators, delay lines, and reflective delay lines with different center/operating frequencies have been widely investigated for gas sensing applications [2]. Selective and sensitive responses towards a chemical/gas are usually achieved by coating a material sensitive to the chemical. The sensing layer modifies the wave velocity and attenuation through changes in its mass, conductivity, or elasticity upon interaction with the exposed chemical thereby allowing to quantify the analyte by measuring the wave parameters [5]. For trace gases, mass-based sensing layers are preferred since the loading of gas species into the sensing layers or their removal can be realized by a fully reversible physical adsorption without the requirement for elevated device temperatures. Recent studies have demonstrated the potential for developing fast, sensitive, and reversible trace gas sensors by integrating engineered nanoporous sensing layers such as metal-organic frameworks (MOFs) and zeolites with SAW transducers [3, 6, 7]. Further improvement in the sensitivity and detection limit are desirable for monitoring gases in many fields such as in natural gas infrastructure, combustion processes, commercial buildings, and industrial manufacturing processes to name a few.

The sensitivity of a SAW sensor can be improved by optimizing either the transducer design/characteristics or the sensing layer properties, or both. It is known that the mass sensitivity of acoustic devices increases with initial frequency. However, most published works have concerned operating frequencies of a few hundred MHz or less. There are several challenges associated with higher frequency devices including the difficulty in patterning electrodes due to the lithographic resolution (SAW velocity in most common piezoelectric crystals is < 4000 m/s [2]), increased noise level, signal loss, and potential secondary effects. For this reason, studies have focused on tuning sensing layer properties such as their gas adsorption capacity and coupling with the acoustic waves [7]. The most recent investigations have begun exploring nanoporous materials such as MOFs and zeolites as sensing layers to improve the sensitivity of the mass-based gas sensors for room temperature applications by leveraging their engineered porosity and physisorption-based interactions with gases [3, 6]. However, achieving ppm-level detection of atmospheric gases, especially small nonpolar and quadrupolar molecules like $CO_2$ and $CH_4$, at ambient conditions is challenging due to their light mass and weak interaction with most sensing materials. In these cases, transducer-level optimization such as development of higher frequency devices

J. Devkota is with National Energy Technology Laboratory, Pittsburgh, PA, 15236, USA and with Leidos Research Support Team, Pittsburgh, PA 15236, USA (e-mail: jagannath.devkota@netl.doe.gov).

D. W. Greve is an emeritus professor at Carnegie Mellon University, Pittsburgh, PA, 15213, USA. He is now with DWG Consulting, Sedona, AZ 86336, USA (e-mail: dw07@andrew.cmu.edu).

T. Hong is with National Energy Technology Laboratory, Pittsburgh, PA, 15236, USA (e-mail: tao.hong@netl.doe.gov).

K.-J. Kim is with National Energy Technology Laboratory, Pittsburgh, PA, 15236, USA and with Leidos Research Support Team, Pittsburgh, PA 15236, USA (e-mail: ki-joong.kim@netl.doe.gov).

P. R. Ohodnicki is with National Energy Technology Laboratory, Pittsburgh, PA, 15236, USA (e-mail: paul.ohodnicki@netl.doe.gov).

becomes critical. In Rayleigh wave-based devices, mass sensitivities vary approximately as the square of the operating frequency suggesting the potential of four times improvement in the sensitivity with doubling the frequency [8-10]. However, increased noise level, and signal losses (both in transducer and films), and potential interferences from secondary effects may create challenges in achieving the expected sensitivity. Some studies have shown that higher frequency may even worsen the detection limit in some cases such as when sensing layers are lossy [11]. Therefore, a thorough investigation on the frequency effect to various factors including the mass sensitivity, the sensing layer characteristics, and the noise level becomes important to understand the overall performance of these sensors with specific designs and sensing layers. Here, we report on the effect of doubling the operating frequency of SAW reflective delay line sensors with zeolitic imidazolate framework-8 (ZIF-8) MOF layers to the sensitivity and detection limit for $CO_2$ and $CH_4$. We also present the successful demonstration of the wireless detection of $CO_2$ and $CH_4$ with the proposed sensor operating at frequencies as high as 860 MHz.

## II. Theory

Finite element simulations were performed using COMSOL 5.4. Simulations were performed first in order to predict the operating frequency of SAW devices with an interdigitated transducer (IDT) having period of 4 and 8 µm. Simulations were then performed to predict the relative mass sensitivity of these devices.

The 2D simulation domains had a length $l$ = 4 or 8 µm and were 2.5$l$ in height. The material parameters were those of Y-Z $LiNbO_3$. This piezoelectric substrate has a large electrochemical coupling factor (4.5%) [2]. Periodic boundary conditions were imposed on left and right boundaries and a fixed boundary condition on the bottom surface. As SAW displacements are greatest at the top surface and decay strongly away from the surface, increasing the height or changing the boundary condition on the bottom surface has negligible effect on the results.

Eigenfrequency simulations were performed to find the frequency corresponding to a surface acoustic wave with wavelength/periodicity of 4 or 8 µm. Fig. 1(a) and (b) show the resulting eigenmodes. The eigenfrequencies were predicted to be 436.38 MHz and 872.76 MHz for 8- and 4-µm structures, respectively. The predicted eigenfrequencies correspond to a SAW velocity of 3491 m/s, consistent with the reported velocity at the free surface of this material [2].

We now turn to the relative gas sensitivity of 4- and 8-µm devices. When used as a sensor, a gas-sensitive layer is deposited on the SAW surface. The sensitive layer considered here is ZIF-8 MOF that adsorbs effectively particular gas species such as $CO_2$ and $CH_4$ [3]. Gas adsorption changes the mass density of the MOF, changing the SAW velocity. For thin films (h << λ) of soft materials [10], the Sauerbrey equation approximates the frequency change to the mass change as,

$$\Delta f = -C f_0^2 \frac{\Delta m}{A} \qquad (1)$$

where $\Delta f$ is the shift in the resonant frequency $f_0$, $C$ is the mass sensitivity coefficient of the piezoelectric substrate (-5.505×10$^{-8}$ m$^2$/s/kg for Y-Z $LiNbO_3$), $A$ is active area, and $\Delta m$ is the loaded mass [10]. However, this equation does not account for possible changes in the wave mode as the frequency is scaled up. Our simulations suggest that this may be a significant effect.

In a previous report [1], we showed that there are changes to the surface wave mode when the sensing layer thickness increases. This phenomenon becomes significant when the sensing layer is close to a quarter transverse wavelength in the sensing layer. As ZIF-8 has a relatively small Young's modulus, this can occur at sensing layer thicknesses used in the experiments reported here. For particular sensing layer thickness, this effect first becomes apparent in 4-µm periodicity devices.

In order to predict the relative sensitivities of the two devices, we performed eigenmode simulations for a domain with a 240 nm thick ZIF-8 layer on the propagation path. Simulations were performed with the nominal mass density ($\rho$) of the ZIF-8 layer and for mass densities increased by up to 1%. As expected, the frequency change is linear in the mass density (not shown). From the change of eigenfrequency we can calculate the relative sensitivity of devices with the same sensing layer thickness as

$$r = \frac{f_{4\mu m}(\rho) - f_{4\mu m}(1.01\cdot\rho)}{f_{8\mu m}(\rho) - f_{8\mu m}(1.01\cdot\rho)} \qquad (2)$$

The results obtained depend on the properties of the ZIF-8 layer. There are two literature reports of the properties of ZIF-8 layers, one concerning single crystal material [12] and a second on (002) texture thin-film material [13]. There is an appreciable range in the reported Young's moduli (2.8 – 4 GPa) and Poisson ratios (ν = 0.33 – 0.54). Further, other reports [14, 15] have noted that ZIF-8 becomes stiffer upon $CH_4$ incorporation. As there is an appreciable range in the reports, and in any case thin film properties may well depend on texture and processing, we have performed simulations for a range of material properties. Additional parameters necessary for the calculation are the relative permittivity ($\varepsilon_r$ = 2.3) and the mass density ($\rho$ = 950 kg/m$^3$) [13].

The results for a range of Young's modulus and Poisson ratio are shown in Fig. 2(a). The results are weakly dependent on Poisson ratio over a range from 0.35 to 0.45. Young's modulus has a large effect, however, with sensitivity ratio substantially greater than 4 predicted for the lowest value and sensitivity ratio close to that predicted by the Sauerbrey equation for values of 6 GPa and above.

The deviation from the factor of 4 from the Sauerbrey equation is a consequence of the change in the character of the surface wave in the 4 µm device when the thickness of the sensing layer approaches a quarter of the transverse wavelength. The transverse wavelength is quite short for the smallest value of the Young's modulus but becomes longer as the Young's modulus increases.

The change to the surface wave mode can be seen both in the surface wave velocity and the energy density of the wave mode. Fig. 2(b) shows the surface wave velocity as a function of ZIF-8 thickness for sensing layer parameters E = 3 GPa and ν = 0.4. As the sensing layer thickness approaches one quarter wavelength of the transverse wave, the surface wave velocity begins to decrease sharply. For larger sensing layer thickness, a second higher-velocity surface wave mode begins to appear. For some range of thicknesses both modes can propagate [1]. The change in character of the surface wave mode is shown in Fig. 3. For a 240 nm MOF sensing layer, there is a substantial kinetic energy density in the sensing layer for E = 3 GPa. At E = 9 GPa the kinetic energy density in the sensing layer is much smaller. As more of the wave kinetic energy is located in the sensing layer for small Young's modulus, it is reasonable that the effect of a mass change in the sensing layer would be accentuated. These effects are not pronounced in 8-μm structure for 240 nm thick ZIF-8 layer.

The higher sensitivity at higher frequencies is attributed to the interaction of a larger fraction of the acoustic energy density with the loaded mass because higher frequency results in localization of the acoustic wave closer to the surface. Consequently, at higher frequencies a given mass change represents a larger fraction of the material moving, resulting in a higher factor of sensitivity increment.

## III. Design, Fabrication, and Characterization

Reflective delay line devices were designed with 8 μm and 4 μm wavelengths consisting of a solid electrode-type bidirectional emitting IDT with an IDT-type reflector on its either side (Fig. 4(a)) and fabricated on Y-Z $LiNbO_3$ by depositing aluminum electrodes of thickness 120 nm and 106 nm, respectively. The number of figure pairs N and the aperture W of the IDT were chosen such that the excited waves have relatively smaller bandwidth ($BW|_{-4dB} \approx 100\%/N$) and lower diffraction [16]. The number of finger pairs N in the reflectors were chosen for stronger reflection with smaller bandwidth [17]. Table 1 summarizes the design parameters of the transducers. Fig. 4(a) is a schematic of the transducers' cross-section view whereas Fig. 4(b) and 4(c) are the SEM images of the representative IDTs of the 8-μm and 4-μm periodicity devices, respectively. The images showed that the electrode widths/spacings in the 8-μm and 4-μm structures are ~2 μm and ~1 μm with acceptable deviations as expected. Even though both device configurations were designed with identical metallization ratio η = 0.5 (ratio of electrode width to the total of the electrode width and spacing), the fabricated 4-μm periodicity structure had a slight deviation from the original design (η ~ 0.5 and ~ 0.7 for 8-μm and 4-μm periodicities, respectively). This deviation is due to resolution issues in the mask development and photolithography processes created by reduced feature sizes of the higher frequency devices. However, the deviation is anticipated to minimally affect the center/operating frequency of the device in the present case since the period remained unchanged. Such deviations are expected to alter the intensities of the generated waves, especially of higher order modes [16].

Parts of the fabricated devices that included the longer delay paths were coated with ZIF-8 layer and the shorter paths were used as the reference to compensate the temperature effect. Fig. 5(a) shows the optical image of a fabricated SAW device with ZIF-8 sensing layer in the longer delay path. As shown, the sensing layer covered most of the delay path as well as the reflector. Fig. 5(b) is the representative scanning electron microscopy (SEM, FEI Quanta 600) image of the surface of ZIF-8 layer. The SEM image shows a uniform and dense film indicating a good quality overlayer. Cross-sectional SEM image shows that the ZIF-8 film is about ~240 nm thick (Fig. 5(c)). The films were coated using solution processable dip-coating in two cycles. Details of the ZIF-8 sensing layer development and characterization can be found elsewhere [3, 18].

The fabricated 8-μm and 4-μm periodicity devices were characterized using a vector network analyzer (Rohde & Schwarz, ZVB4) over wide frequency spans 415 – 445 MHz and 840 – 875 MHz, respectively. The measured frequency spans were wide enough to capture most of the dispersion characteristics. The scans were then transformed into the time domain using an inverse Fourier transform. Fig. 5(d) is the inverse fast Fourier transform of the reflection characteristics of a representative 4-μm device with and without ZIF-8 sensing layer on its longer delay path. For uncoated 8-μm and 4-μm periodicity devices, the acoustic reflection peaks $R_1$ and $R_2$ occurred at $t_1$ = 1.35 μs and $t_2$ = 1.72 μs and at $t_1$ = 1.18 μs and $t_2$ = 1.63 μs, respectively. These time delays are consistent with the expected delays for the designed delay paths on Y-Z $LiNbO_3$. In Fig. 5(d), the right peak inside the dotted circle represents the delay and attenuation caused by the sensing layer on the 4-μm device whereas the left peak represents the reflection from the reference delay path. The measured delay and attenuation of the waves due to the film in the delay path were about 0.06 μs and 8.10 dB, respectively.

Information about the exposed gases to these sensors can be obtained by monitoring the delay or the attenuation in real time and correlating with the added mass to the film upon exposure. In principle, sensors with higher center frequency and thicker sensing layers are expected to possess higher sensitivity in terms of the absolute frequency shift as discussed and explained above. However, the dependency of the attenuation and noise level on the center frequency and sensing layer characteristics including the thickness and roughness may limit the sensor responses. Thicker and/or rougher films can potentially weaken the acoustic signals through attenuation (due to absorption of larger amount of energy) and increase in baseline drift. This effect becomes severe in higher frequency devices [19]. There is always a trade-off between the sensor's sensitivity and its characteristics including the transducer geometry, center frequency, and the film quality and thickness. The experimental investigation focuses on comparing the sensitivity and detection limit of the devices with two unique initial frequencies and nominally identical sensing layers with the thickness below the critical values suggested by our simulation.

TABLE I
SAW Devices Design Parameters

| Parameters | Low Frequency Device | High Frequency Device |
|---|---|---|
| Wavelength (λ) | 8 μm | 4 μm |
| Electrode Width and Spacing | 2 μm | 1 μm |
| IDT Type/Fingers | Solid/50 pairs | Solid/50 pairs |
| Aperture | 100 λ | 200 λ |
| Reflectors (Fingers, Distance) | $R_1$ (30 pairs, 2.35 mm) $R_2$ (50 pairs, 3.05 mm) | $R_1$ (30 pairs, 2.05 mm) $R_2$ (50 pairs, 2.88 mm) |

## IV. Gas Testing

### A. Wired Measurements

The gas sensing tests were performed in a 100 mL chamber for a desired gas mixture (various concentrations of $CO_2$ or $CH_4$ in $N_2$) flowed at a rate of 100 mL/min. The phase associated with the time delay ($\Delta\phi = 2\pi f_0 \Delta t$; see Fig. 5(d) for example) was measured in wired or wireless mode using a transient radar interrogator [20]. Fig. 6(a) shows a schematic of the gas testing system and its inset shows an optical image of a wireless SAW sensor inside the test chamber. RF pulses of 10 dBm power and frequency $f_0$ (430 MHz for 8-µm devices and 860 MHz for 4-µm devices) were excited, the phases of the excited and reflected pulses were recorded *via* I/Q method, and the differences were calculated [20]. The measured phase delay of the reference path was subtracted from that of the sensing layer-coated path ($\Delta\phi = \Delta\phi_2 - (L_2/L_1)\Delta\phi_1$, where $L_1$ and $L_2$, are the lengths of uncoated and coated paths, respectively) to compensate the temperature effect [21]. Fig. 6(b) and (c) depict the wired measurement of real time phase responses of 8-µm and 4-µm sensors, respectively to various concentrations of $CO_2$ at ambient condition. As shown, the phase shifted from its $N_2$ baseline when $CO_2$ was exposed. Under linear approximation, the phase is proportional to the wave velocity so that the fractional change in the phase delay is equivalent to that of the wave velocity which, in this case is due to the mass loading into the film when the test gas is adsorbed. A larger shift was observed for higher concentrations due to adsorption of larger amount of the gas to cause a higher mass loading. Also, the higher frequency (4-µm periodicity) device had a higher phase shift for a given gas concentration. For instance, exposure of pure $CO_2$ to the 8-µm and 4-µm sensors shifted their phases by 1.11 rad and 4.58 rad, respectively, from their respective $N_2$ baseline phases.

To quantify the sensitivities of the two device configurations, we plotted the phase shifts against the $CO_2$ and $CH_4$ gases concentrations which showed a linear response to the concentrations (Fig. 7). The slopes of these linear curves were used as the sensor sensitivities to the target gases. For $CO_2$ (Fig. 7(a)), the 4-µm sensor was about 4.1 times more sensitive (sensitivity, $\eta$ = 2.64 deg/vol-%) than the 8-µm sensor ($\eta$ = 0.63 deg/vol-%). Similarly, the $CH_4$ sensitivity (Fig. 7(b)) of the 8-µm and 4-µm devices were obtained to be 0.03 deg /vol-% and 0.14 deg /vol-%, respectively, which results to a sensitivity enhancement by a factor of 4.2, similar to that obtained for $CO_2$. The phase shift for $CH_4$ is much lower compared to that for $CO_2$ for two reasons - (i) smaller amount of the gas uptake in the film due to larger kinetic molecular diameter and lower polarizability and (ii) smaller molecular weight (closer to that of $N_2$). The sensing mechanism of a thin ZIF-8 film-coated SAW reflective delay line sensors for a variety of gases and gas concentrations has been reported in our previous study [3]. It is important to note that the difference in the active sensing area of the sensors may also potentially alter their sensitivities. However, the current configurations have identical apertures that makes the ratio of the phase shifts ($\Delta\phi_r$) independent of the active area to the first order ($\Delta\phi/\phi = -Cf\Delta m/A; \phi = 2\pi fL/v$, $L$ is delay length) for full coverage of the delay paths.

The experimental observation of a sensitivity ratio almost equal to 4 (in agreement with the Sauerbrey approximation) suggests that the properties of ZIF-8 film coated in this study are somewhat different from those reported in the literature. This could be a consequence of stiffening due to gas incorporation in the course of our measurements. Alternatively, the thin film preparation method used here may result in stiffer films than those reported in the literature. Our simulations, and in particular the appearance of changes to the surface wave mode, suggest that soft sensing layers may exhibit enhanced mass sensitivity at higher frequencies under some circumstances.

An important factor that can limit the performance of the higher frequency sensors is the noise level that increases with frequency. To estimate the noise levels of the MOF-coated sensors, we calculated the root mean squared (*RMS*) error as $RMS_{error} = ((\Sigma x_i - x_{avg})^2/n)^{1/2}$ of the repeated phase measurements recorded for constant $N_2$ flow (30 minutes) and smoothed (using FFT filter in OriginLab 2019) data. The RMS noises for the 8 µm and 4 µm devices were obtained to be 0.0033 and 0.0058 rad, respectively. Using these noises and the sensitivity extracted as the slopes of the curves in Fig. 7, we estimated the detection levels (= 3×noise level/sensitivity) for the gases [22]. The $CO_2$ detection limits for the 8-µm and 4-µm sensors were estimated to be 0.91 vol-% and 0.38 vol-%, respectively. Similarly, the $CH_4$ detection limits for the 8-µm and 4-µm sensors were estimated to be 16.60 vol-% and 7.01 vol-%, respectively.

### B. Wireless Measurements

Wireless and passive SAW sensors are suffered by several noise sources including electromagnetic and chemical interferences and signal losses (e.g. path loss between antennas, insertion loss, and losses in the sensor device) that create challenges in reliable sensing and operating from a longer distance [23]. In case of chemical sensors such as the one presented here, additional challenges are created by chemical interferences and losses in the sensing layers. The losses are more pronounced in the higher frequency devices that make the wireless measurement more challenging. Therefore, much attention is paid on reducing the losses and minimizing the interferences when developing wireless sensors through an optimized transducer and antenna design, use of low-loss materials (transducer and sensing layer), and development of matching circuit to match the antenna and transducer impedances [23]. We, however, report here a wireless detection of the test gases at 860 MHz using a custom fabricated half-wave dipole antenna from 5 cm without using any matching circuit or optimizing the transducer and antenna designs. First, we fabricated the antennas using a 24 AWG copper wire with PVC jacket and characterized using R&S

ZVB4 VNA. The antennas were then integrated with the 4-μm structure SAW devices (Fig. 6(a), inset) and measured the sensor responses similarly as explained above in wired mode measurements. Fig. 8(a) and (b) show the wirelessly measured real time sensor responses to various concentrations of $CO_2$ and $CH_4$, respectively. The wireless measurements for the gases were consistent with the wired measurements (see Fig. 6(c) and Fig 8(a) for $CO_2$). Fig 8(b) shows the ability of the sensor to detect as low as 10 v/v% $CH_4$. In near future, we plan to report on the gas testing from an optimized interrogation distance through the use of an antenna well-matched to the SAW sensors either by design or by using a matching circuit. Present observations are of high importance in continued development of sensitive wireless and passive $CO_2$ and $CH_4$ sensors for use in energy infrastructure monitoring applications.

## V. Conclusion

We investigated the effect of operating frequency on the sensing response of a mass-based SAW sensor coated with ZIF-8 MOF to $CO_2$ and $CH_4$ gases at ambient condition. The experimentally observed sensitivity of the 240 nm thick ZIF-8 coated 4 μm periodicity sensor was about four times higher than that of 8 μm periodicity sensor which agrees with the Sauerbrey approximation and slightly lower than the FEM prediction. The simulation suggested that the appearance of higher frequency surface wave modes and changes in the acoustic energy in soft sensing layers may alter the sensitivity of the sensors. We also successfully demonstrated the wireless detection and monitoring of the gases using 4 μm (860 MHz) sensor. These results are promising towards developing wireless and passive gas sensors for a range of applications including natural gas leak detection, carbon capture and sequestration, other fossil fuel industries, and in large buildings.

## Acknowledgment

This work was performed in support of the US Department of Energy's Fossil Energy Crosscutting Technology Research Program. The Research was executed through the NETL Research and Innovation Center's Natural Gas Transmission and Delivery FWP. Research performed by Leidos Research Support Team staff was conducted under the RSS contract 89243318CFE000003.

## Disclaimer

## References

[1] D. W. Greve, J. A. Devkota, and P. Ohodnicki, "Wireless $CO_2$ SAW Sensors with a Nanoporous ZIF-8 Sensing Layer," in *2018 IEEE International Ultrasonics Symposium (IUS)*, 2018, pp. 1-4.

[2] J. Devkota, P. R. Ohodnicki, and D. W. Greve, "SAW Sensors for Chemical Vapors and Gases," *Sensors,* vol. 17, Apr 2017.

[3] J. Devkota, K.-J. Kim, P. Ohodnicki, J. T. Culp, D. Greve, and J. W. Lekse, "Zeolitic imidazolate framework-coated acoustic sensors for room temperature detection of carbon dioxide and methane," *Nanoscale,* 2018.

[4] V. P. Plessky and L. M. Reindl, "Review on SAW RFID tags," *IEEE Trans Ultrason Ferroelectr Freq Control,* vol. 57, pp. 654-68, Mar 2010.

[5] A. J. Ricco and S. J. Martin, "Thin Metal-Film Characterization and Chemical Sensors - Monitoring Electronic Conductivity, Mass Loading and Mechanical-Properties with Surface Acoustic-Wave Devices," *Thin Solid Films,* vol. 206, pp. 94-101, Dec 10 1991.

[6] B. Paschke, A. Wixforth, D. Denysenko, and D. Volkmer, "Fast Surface Acoustic Wave-Based Sensors to Investigate the Kinetics of Gas Uptake in Ultra-Microporous Frameworks," *ACS Sensors,* vol. 2, pp. 740-747, 2017/06/23 2017.

[7] A. Afzal, N. Iqbal, A. Mujahid, and R. Schirhagl, "Advanced vapor recognition materials for selective and fast responsive surface acoustic wave sensors: A review," *Analytica Chimica Acta,* vol. 787, pp. 36-49, Jul 17 2013.

[8] F. L. Dickert, P. Forth, W.-E. Bulst, G. Fischerauer, and U. Knauer, "SAW devices-sensitivity enhancement in going from 80 MHz to 1 GHz," *Sensors and Actuators B: Chemical,* vol. 46, pp. 120-125, 2/15/ 1998.

[9] A. J. Ricco and S. J. Martin, "Multiple-frequency SAW devices for chemical sensing and materials characterization," *Sensors and Actuators B: Chemical,* vol. 10, pp. 123-131, 1993/01/01/ 1993.

[10] H. Wohltjen, "Mechanism of Operation and Design Considerations for Surface Acoustic-Wave Device Vapor Sensors," *Sensors and Actuators,* vol. 5, pp. 307-325, 1984.

[11] J. W. Grate and M. Klusty, "Surface acoustic wave vapor sensors based on resonator devices," *Analytical Chemistry,* vol. 63, pp. 1719-1727, 1991/09/01 1991.

[12] J. C. Tan, B. Civalleri, C. C. Lin, L. Valenzano, R. Galvelis, P. F. Chen, *et al.*, "Exceptionally Low Shear Modulus in a Prototypical Imidazole-Based Metal-Organic Framework," *Physical Review Letters,* vol. 108, Feb 29 2012.

[13] S. Eslava, L. Zhang, S. Esconjauregui, J. Yang, K. Vanstreels, M. R. Baklanov, *et al.*, "Metal-Organic Framework ZIF-8 Films As Low-κ Dielectrics in Microelectronics," *Chemistry of Materials,* vol. 25, pp. 27-33, 2013/01/08 2013.

[14] Z. Su, Y.-R. Miao, S.-M. Mao, G.-H. Zhang, S. Dillon, J. T. Miller, *et al.*, "Compression-Induced Deformation of Individual Metal–Organic Framework Microcrystals," *Journal of the American Chemical Society,* vol. 137, pp. 1750-1753, 2015/02/11 2015.

[15] A. U. Ortiz, A. Boutin, A. H. Fuchs, and F.-X. Coudert, "Investigating the Pressure-Induced Amorphization of Zeolitic Imidazolate Framework ZIF-8: Mechanical Instability Due to Shear Mode Softening," *The Journal of Physical Chemistry Letters,* vol. 4, pp. 1861-1865, 2013/06/06 2013.

[16] C. Campbell, *Surface Acoustic Wave Devices for Mobile and Wireless Communications, Four-Volume Set*: Academic press, 1998.

[17] J. Devkota, P. R. Ohodnicki, J. A. Gustafson, C. E. Wilmer, and D. W. Greve, "Designing a SAW Sensor Array with MOF Sensing Layers for Carbon Dioxide and Methane," in *2019 Joint Conference of the IEEE International Frequency Control Symposium and European Frequency and Time Forum (EFTF/IFC)*, 2019, pp. 1-4.

[18] K.-J. Kim, P. Lu, J. T. Culp, and P. R. Ohodnicki, "Metal–Organic Framework Thin Film Coated Optical Fiber Sensors: A Novel Waveguide-Based Chemical Sensing Platform," *ACS Sensors,* vol. 3, pp. 386-394, 2018/02/23 2018.

[19] S. Fujii, T. Odawara, H. Yamada, T. Omori, K. Hashimoto, H. Torii*, et al.*, "Low propagation loss in a one-port SAW resonator fabricated on single-crystal diamond for super-high-frequency applications," *IEEE Transactions on Ultrasonics, Ferroelectrics, and Frequency Control,* vol. 60, pp. 986-992, 2013.

[20] T. L. Chin, P. Zheng, D. W. Greve, L. Cao, and I. J. Oppenheim, "Flexible instrumentation for wireless SAW," in *2010 IEEE International Ultrasonics Symposium*, 2010, pp. 261-264.

[21] M. Jungwirth, H. Scherr, and R. Weigel, "Micromechanical precision pressure sensor incorporating SAW delay lines," *Acta Mechanica,* vol. 158, pp. 227-252, 2002/09/01 2002.

[22] D. S. Ballantine Jr, S. J. Martin, A. J. Ricco, G. C. Frye, H. Wohltjen, R. M. White*, et al.*, "Chapter 5 - Chemical and Biological Sensors," in *Acoustic Wave Sensors*, ed Burlington: Academic Press, 1997, pp. 222-330.

[23] A. Pohl, "A review of wireless SAW sensors," *IEEE Transactions on Ultrasonics, Ferroelectrics, and Frequency Control,* vol. 47, pp. 317-332, 2000.

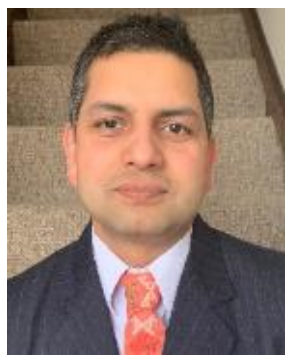

**Jagannath Devkota** received the Ph.D. degree in applied physics from the University of South Florida in 2015. He is a Research Scientist with Leidos Research Support Team, contractor for US-DOE, National Energy Technology Laboratory (NETL) where he was an ORISE Postdoctoral Researcher for 2016 – 2017 and AECOM Research and Development Scientist for 2017 - 2019. He served as a Postdoctoral Research and Teaching Associate at the University of Georgia for a year before joining NETL. His primary research interests are in RF and microwave electronic devices and functional materials for sensing, power electronics, and electromagnetic applications. He has published 32+ research articles, a book chapter, and has 5+ patents under review.

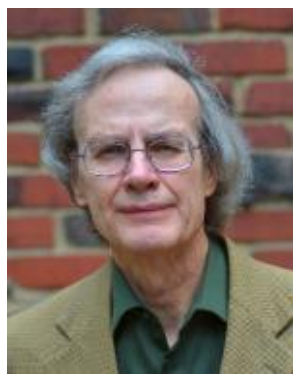

**David W. Greve** David W. Greve received the Ph.D. degree in Electrical Engineering from Lehigh University in 1980. He worked at Philips Research Laboratories, Sunnyvale before joining the (then) Electrical Engineering department at Carnegie Mellon University. He became Professor of Electrical and Computer Engineering at Carnegie Mellon in 1990 and Emeritus Professor in 2016. He presently is Principal at DWGreve Consulting, Sedona, Arizona.

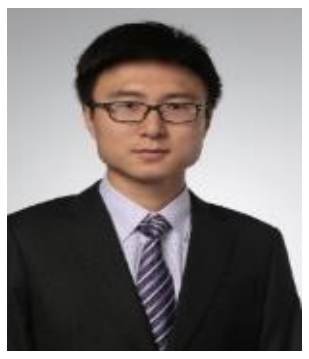

**Tao Hong** received his B.S. degree in Chemistry from Hunan University in 2012 and Ph.D. degree in Polymer Chemistry from the University of Tennessee-Knoxville in 2017. He is now an ORISE postdoctoral fellow of the National Energy Technology Laboratory in Pittsburgh, Pennsylvania. His primary research interests are focused on development and fundamental study of polymeric materials for sensors, thin films and gas separation applications.

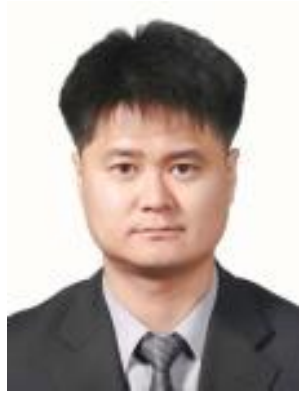

**Ki-Joong Kim** received his PhD degree in chemical engineering from Sunchon National University, South Korea, in 2009. He is currently a research scientist with Leidos Research Support Team (LRST) working as an on-site contractor at National Energy Technology Laboratory (NETL), US DOE. His research focuses on the development of chemical sensing materials and their integration technology as thin films on sensor devices for fossil energy applications. He has more than 60 peer-reviewed journal publications and 4 issued US patents. His work has been cited in scientific publications more than 1000 times.

**Paul R. Ohodnicki, Jr.** graduated from the University of Pittsburgh in 2005 with a B.Phil. in engineering physics and a B.A. in economics and subsequently earned his M.S. (2006) and Ph.D. (2008) in materials science and engineering from Carnegie Mellon University.

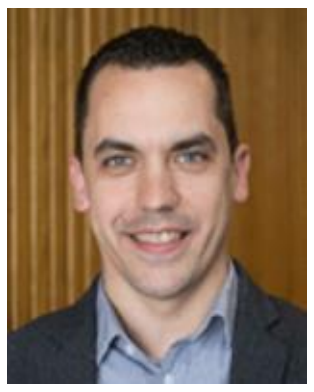

He is currently an associate professor in the Mechanical Engineering and Materials Science department at the University of Pittsburgh and the associate coordinator of the Engineering Science program. Prior to his current role, he was a materials scientist and technical portfolio lead in the Functional Materials Team of the Materials Engineering & Manufacturing Directorate of the National Energy Technology Laboratory. During his time with the laboratory, he was responsible for overseeing projects spanning sensing and power electronics, with emphasis on advanced devices and enabling functional materials for photonic and wireless sensing as well as power magnetics component and materials design. He spent two years as a visiting research scientist at PPG Industries from 2008 to 2010. He has published more than 130 technical publications and holds more than 10 patents, with more than 15 additional patents under review. He also is the recipient of a number of awards and recognitions, including the Federal Employee Rookie of the Year Award (2012), Presidential Early Career Award in Science and Engineering (2016), and the Advanced Manufacturing and Materials Innovation Category Award for the Carnegie Science Center (2012, 2017, 2019). In 2017, he was a nominee for the Samuel J. Heyman service to America Medal.

## Figure Captions

Fig. 1. Eigenmodes for a surface wave with wavelength of 8 µm (left) and 4 µm (right). Color indicates the relative vertical displacements, with blue negative (low) and red positive (high).

Fig. 2. (a) Mass sensitivity ratio as defined in Eq. (2) for 4 and 8 µm devices with the same sensing layer thickness computed as a function of Young's modulus and Poisson ratio. (b) Velocity of two surface wave modes as a function of ZIF-8 thickness. The simulations are for $E = 3$ GPa and $\nu = 0.4$.

Fig. 3. Surface plot of the kinetic energy density for two different values of the Young's modulus. The wave periodicity is 4 µm and the sensing layer thickness 240 nm.

Fig. 4 Schematic of cross-section view of the proposed reflective delay lines (a) and SEM images of emitting IDTs of 8-µm (b) and 4-µm (c) periodicity devices.

Fig. 5. (a) Optical image of a fabricated 8-µm periodicity device with ZIF-8 sensing layer, (b) surface and (c) cross-sectional SEM images of the ZIF-8 film on a device, exhibiting an average thickness of ~240 nm, (d) inverse fast Fourier transformed data of a 4-µm periodicity device with and without ZIF-8 sensing layer.

Fig. 6. Experimental setup for gas testing (a) and real time phase responses of 240 nm thick ZIF-8 coated 8-µm (b) and 4-µm (c) periodicity devices to various concentrations of $CO_2$ measured at 430 MHz and 860 MHz, respectively in wired mode. Inset of (a) is an optical image of a wireless SAW sensor inside the gas chamber.

Fig. 7 Measured phase shifts (dots) of 8-µm and 4-µm periodicity SAW sensors and their linear fits (lines) for various concentrations of $CO_2$ (a) and $CH_4$ (b).

Fig. 8. Real time phase response of a wireless 4-um periodicity SAW sensor with 240 nm ZIF-8 layer to $CO_2$ (a) and $CH_4$ (b) measured at 860 MHz.

## Table Caption

Table 1. SAW devices design parameters.

**Table 1**

| Parameters | Low Frequency Device | High Frequency Device |
|---|---|---|
| Wavelength (λ) | 8 μm | 4 μm |
| Electrode Width and Spacing | 2 μm | 1 μm |
| IDT Type/Fingers | Solid/50 pairs | Solid/50 pairs |
| Aperture | 100 λ | 200 λ |
| Reflectors (Fingers, Distance) | $R_1$ (30 pairs, 2.35 mm)<br>$R_2$ (50 pairs, 3.05 mm) | $R_1$ (30 pairs, 2.05 mm)<br>$R_2$ (50 pairs, 2.88 mm) |

**Abstract Figure**

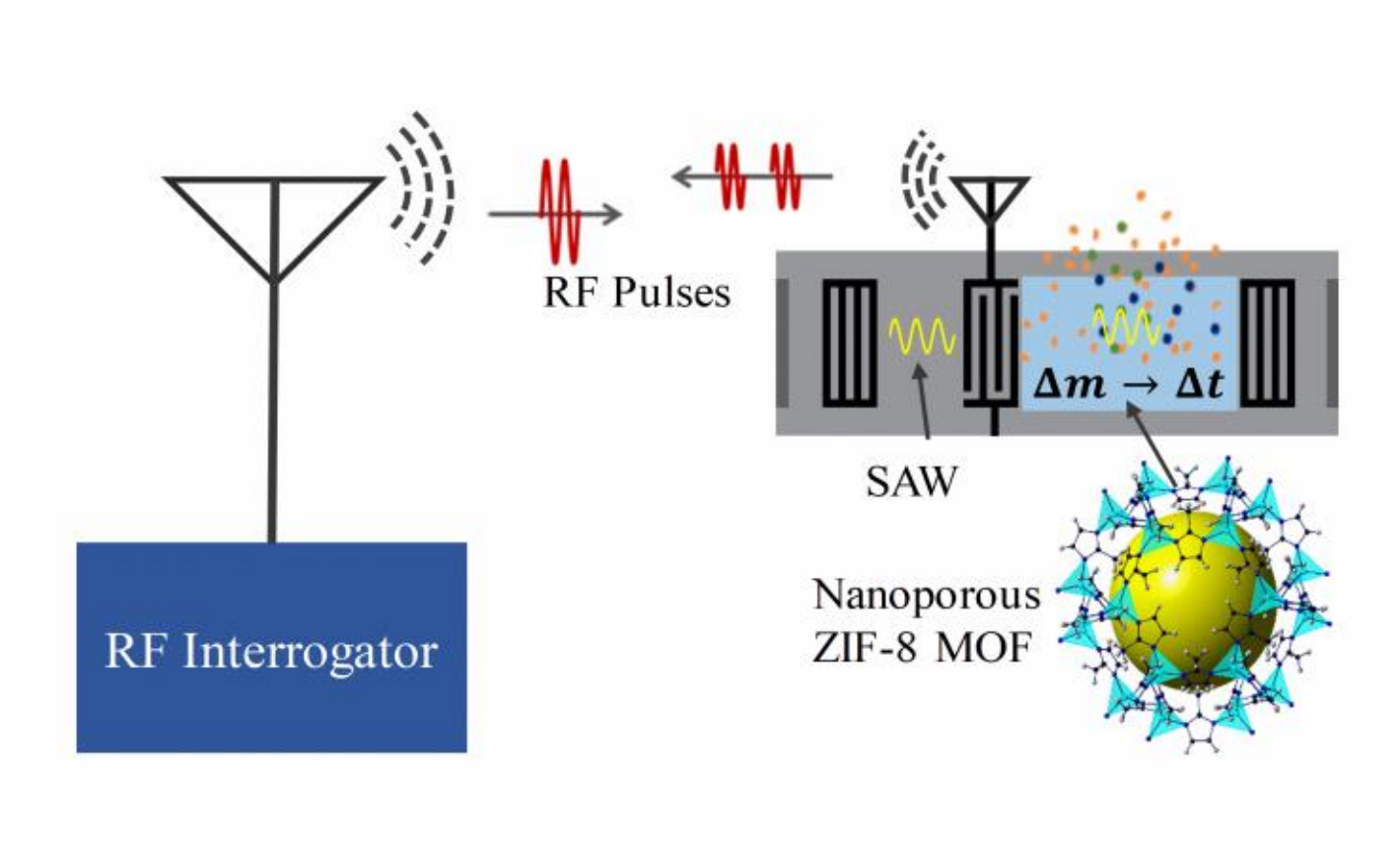

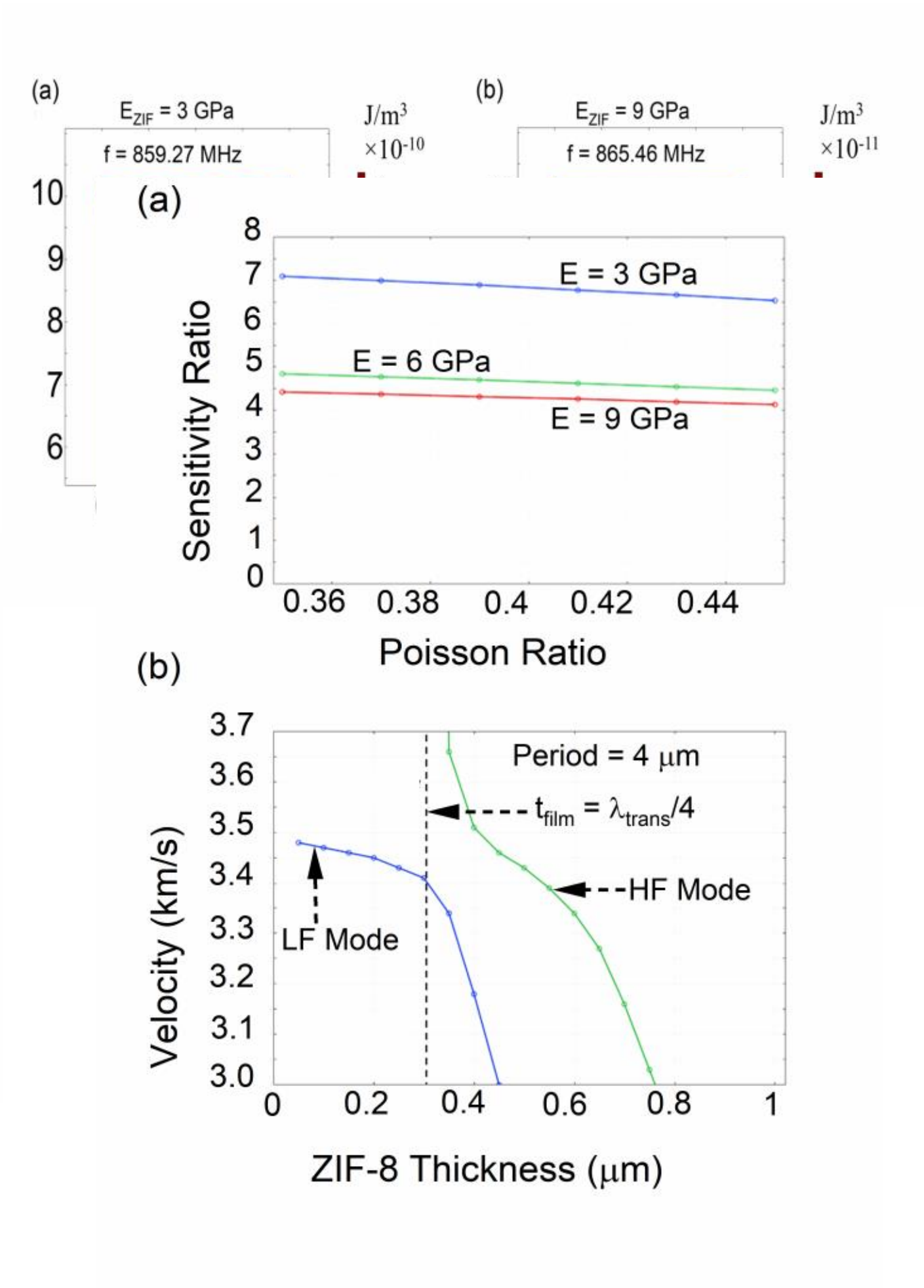

**Fig. 1.**

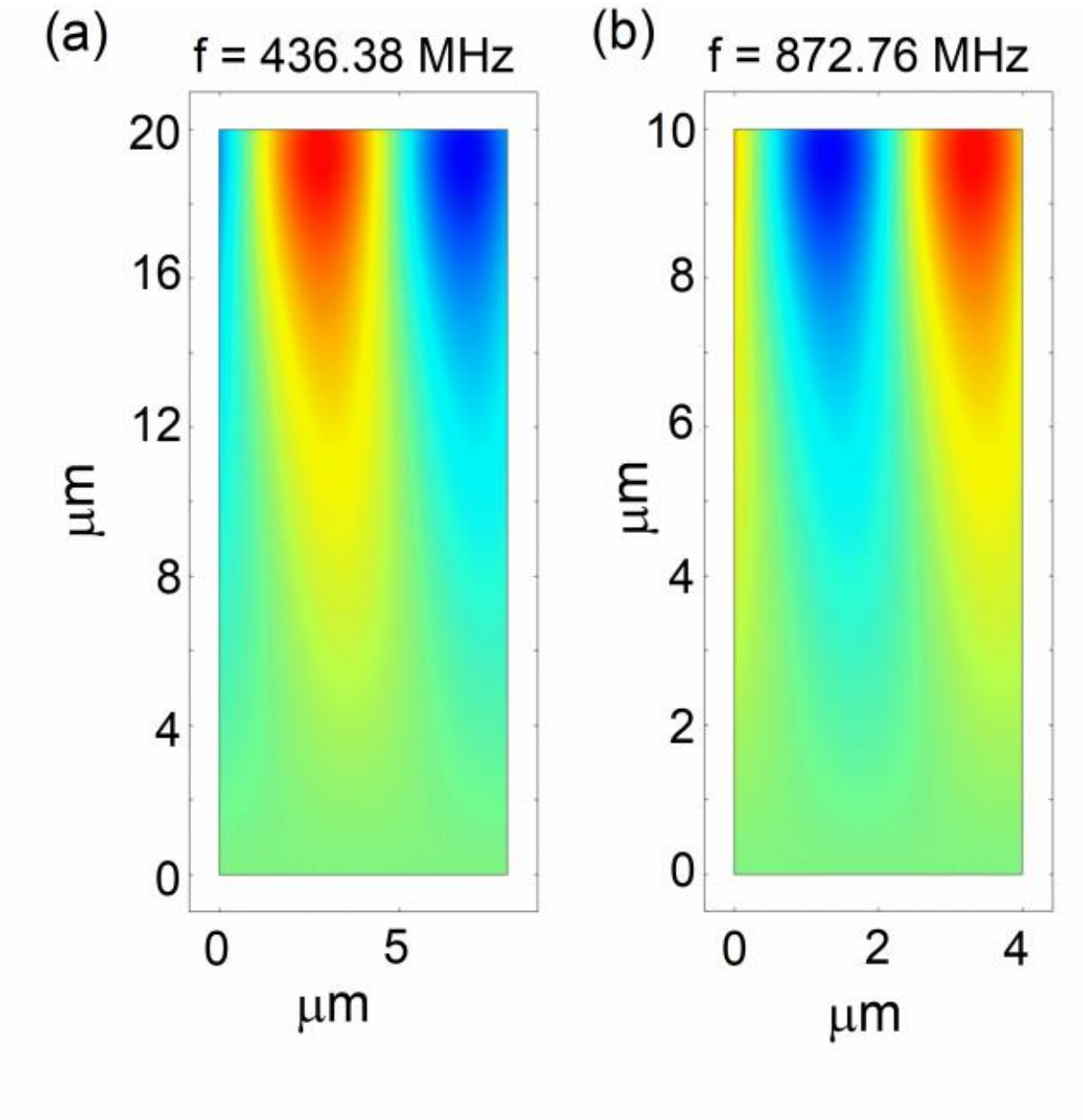

**Fig. 2.**

**Fig. 3.**

**Fig. 4.**

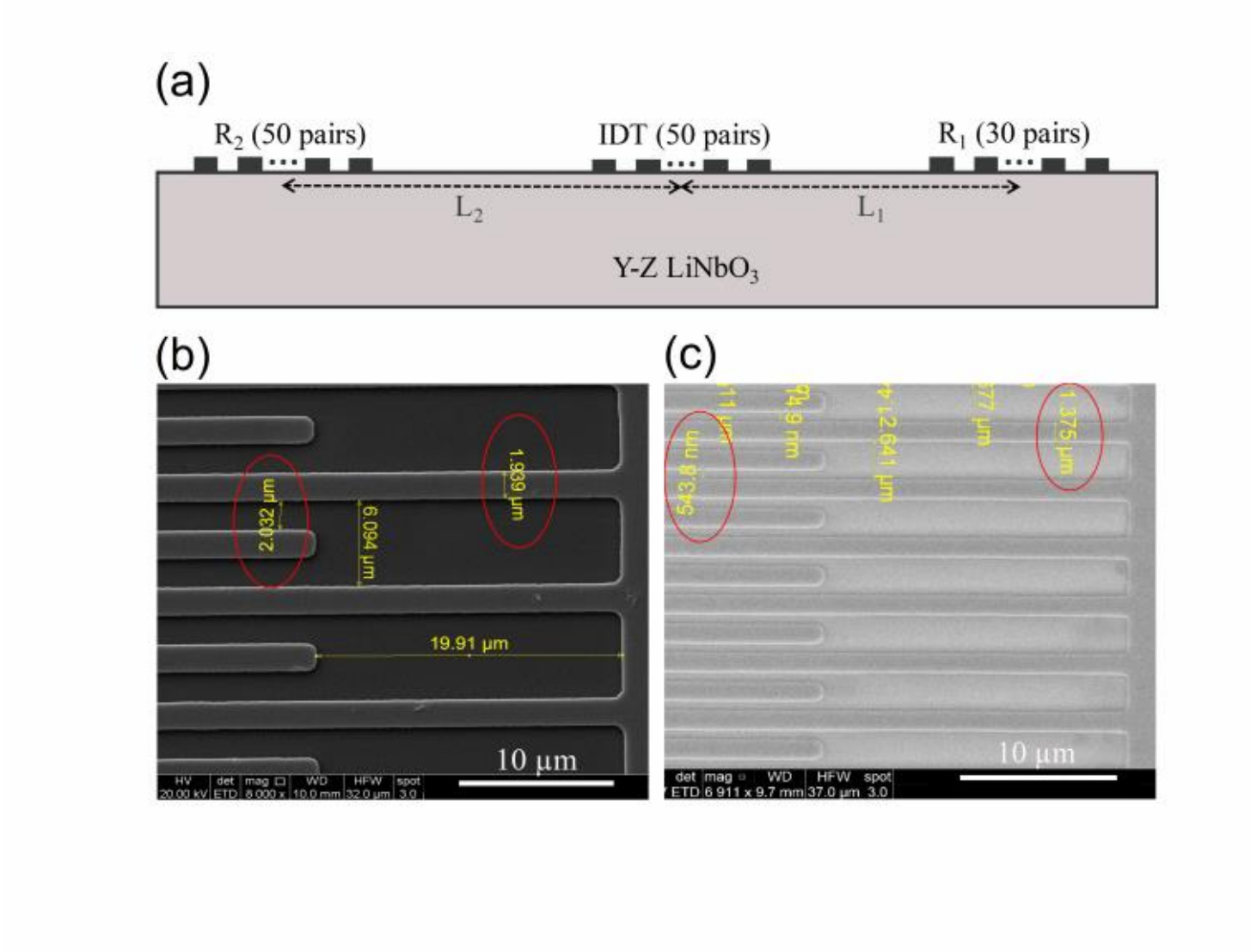

**Fig. 5.**

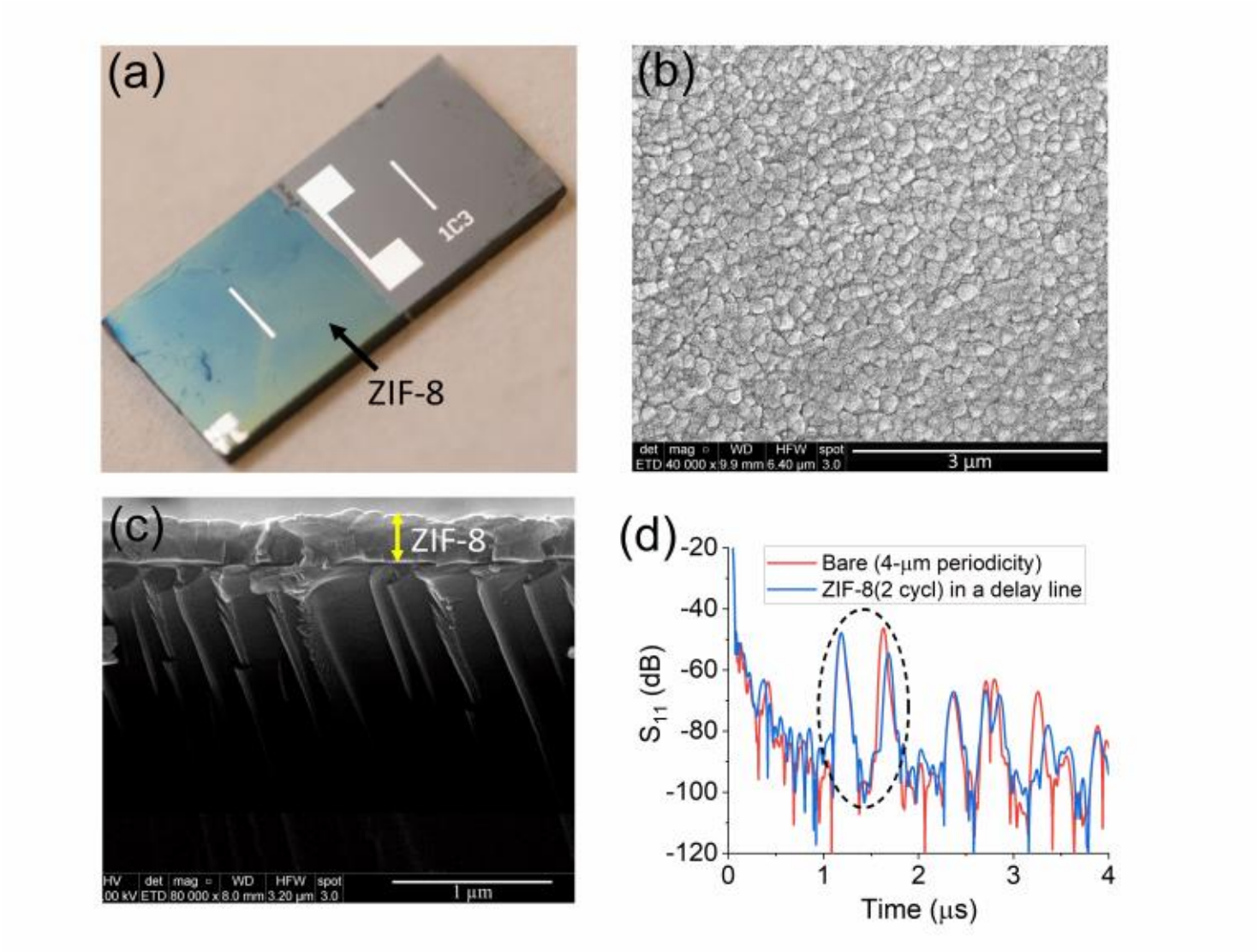

**Fig. 6.**

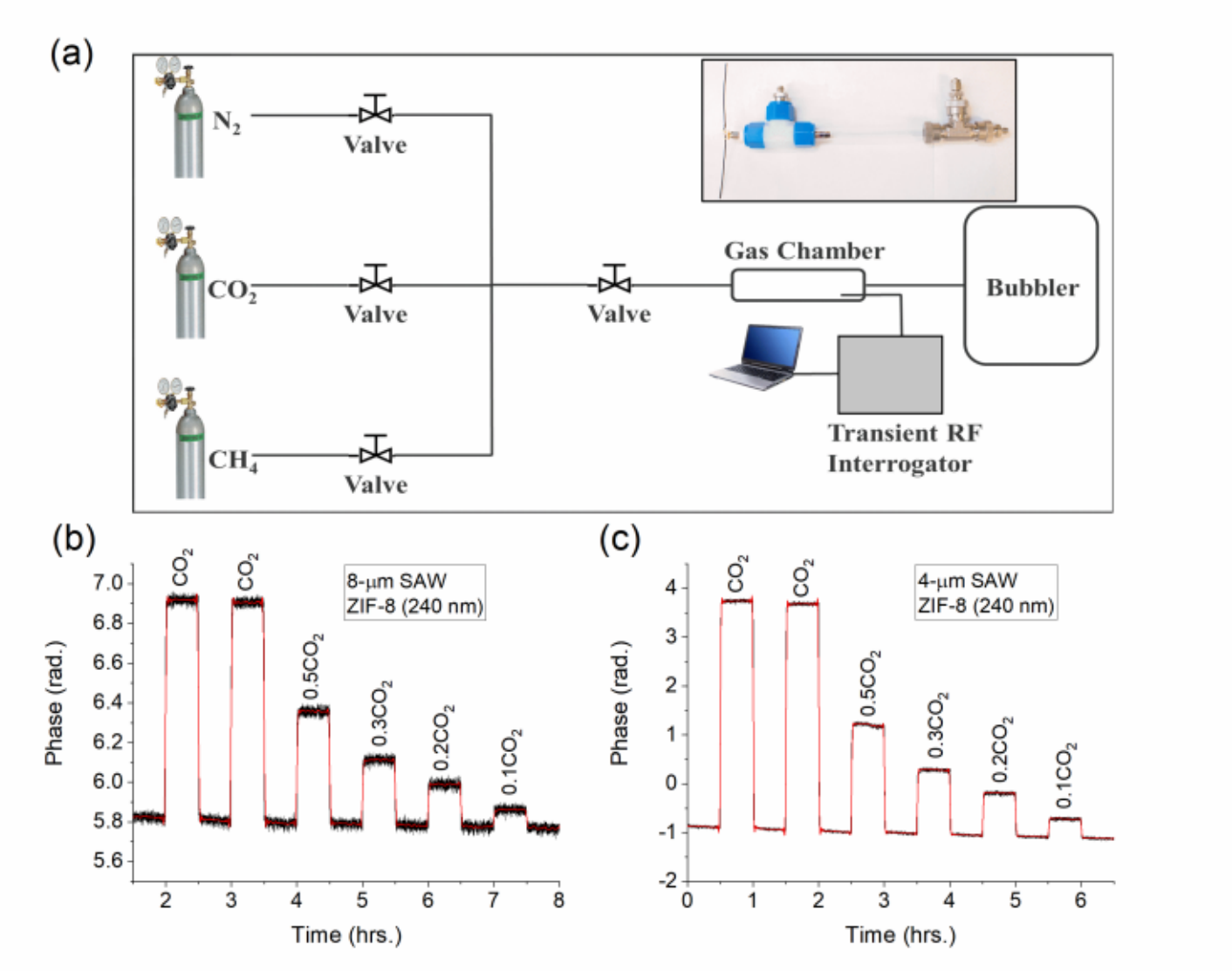

**Fig. 7.**

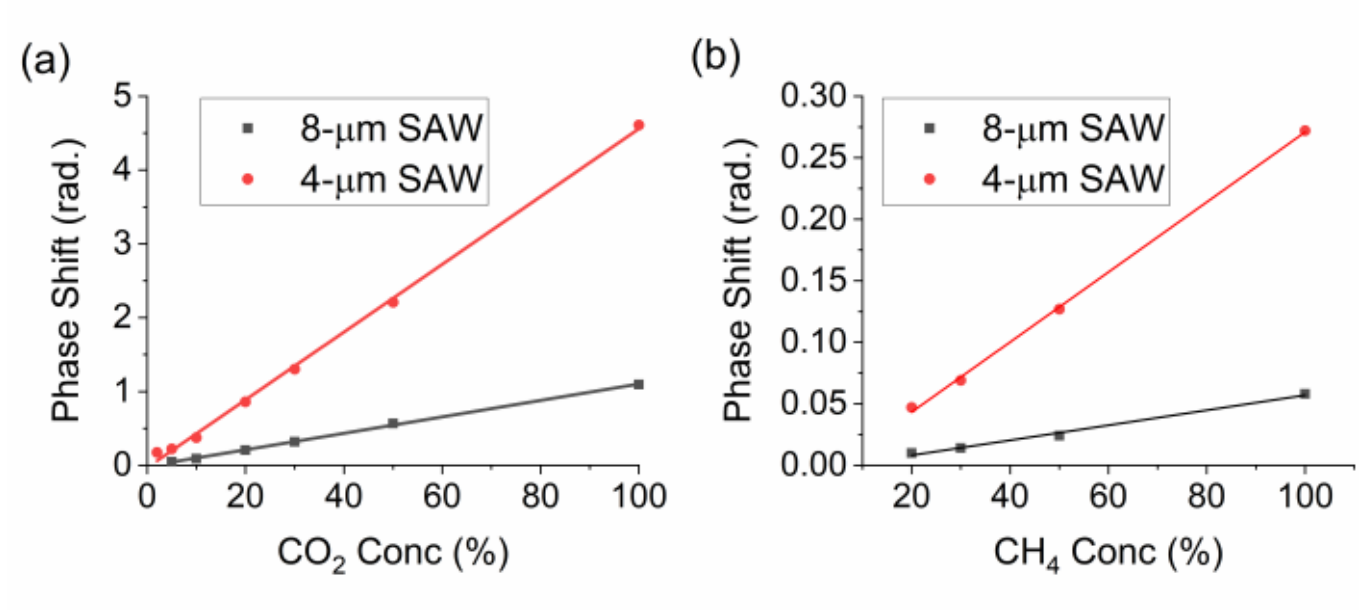

**Fig. 8.**

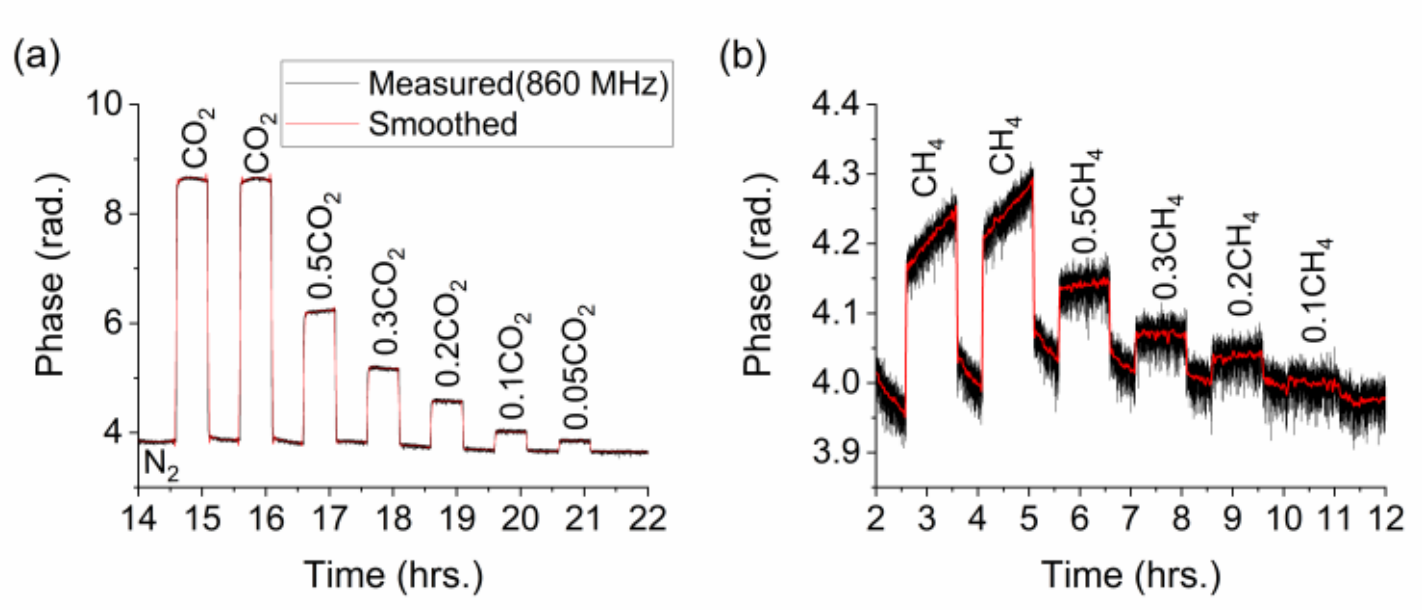